\documentclass[%
 reprint,
 amsmath,amssymb,
 aps,
floatfix,
]{revtex4-1}

\usepackage{graphicx}
\usepackage{dcolumn}
\usepackage{bm}
\usepackage[font=footnotesize,justification=raggedright]{caption} 
\usepackage{amsmath}
\usepackage{amssymb}
\usepackage{color}

\usepackage [english]{babel}
\usepackage [autostyle, english = american]{csquotes}
\MakeOuterQuote{"}
\begin{document}
\title{A simple random matrix model for the vibrational spectrum of jammed packings}
\author{E. Stanifer, P.K. Morse, A.A. Middleton, M.L. Manning}
\affiliation{Syracuse University Department of Physics}
\begin{abstract}
To better understand the surprising low-frequency vibrational modes in structural glasses, we study the spectra of a large ensemble of sparse random matrices where disorder is controlled by the distribution of bond weights and network coordination. We find $D(\omega)$ has three regimes: a very-low frequency regime that can be predicted analytically using extremal statistics, an intermediate regime with quasi-localized modes, and a plateau with $D(\omega) \sim \omega^0$. In the special case of uniform bond weights, the intermediate regime displays $D(\omega) \sim \omega^4$, independent of network coordination and system size, just as recently discovered in simulations of structural glasses.
\end{abstract}
\maketitle

The vibrational spectra of disordered glassy materials exhibit universal features. Although these features govern the mechanical response and provide insight into mechanisms for material failure, their origin remains poorly understood.

Perhaps the most well-studied feature of the density of vibrational states $D(\omega)$ is the boson peak, which is an excess of vibrational modes above the Debye prediction, $D(\omega)\propto \omega^{d-1}$~\citep{Manning2011,OhernAPS,DeGiuli2014}. In jammed packings the frequency at which the peak occurs, $\omega^*$, scales linearly with the average excess number of contacts $\delta z$ above the isostatic point where the number of constraints equals the degrees of freedom~\citep{Wyart2005,Silbert2005,OhernAPS}. Additionally, the eigenvector statistics of modes in the boson peak follow a universal distribution~\citep{Manning2015}.

Recently, another universal feature has been identified in simulations of low-dimensional jammed systems: $D(\omega)\sim\omega^4$ below $\omega^*$~\citep{Lerner2016,Mizuno2017,Kapteijns2018}, which deviates from recent mean-field calculations for the spectra in infinite dimensions that predict $D(\omega)\sim\omega^2$~\citep{Charbonneau2016,Parisi2010}. This interesting behavior has also been found in Heisenberg spin glass systems~\citep{BaityJesi2015}. Understanding this regime is important, as the vibrational modes are quasilocalized and help govern flow and failure in disordered solids~\citep{Manning2011,Wijtmans2017,Tanguy2010,Tsamados2008,Ashton2009,Brito2007,BaityJesi2015}. 


Given the success of random matrix theory in predicting universal features in other physical systems~\citep{mehta2004},
it is natural to wonder if a random matrix model may also explain the $\omega^4$ scaling in jammed packings. Other features, including the boson peak, have already been understood in terms of Euclidean random matrices, which are dynamical matrices for a set of points that are randomly and uniformly distributed in space~\citep{Parisi2002}.

Although there are generic arguments that the global minima of random functions should have a spectrum that scales as $\omega^4$~\citep{Gurarie2003}, we would like to construct a random matrix model to provide insight into how features of the $\omega^4$ region, such as the prefactor, or the location of the scaling regime, change with parameters such as the excess coordination $\delta z$. Such an understanding is important for predicting how material preparation protocols alter the mechanical response of glassy materials.


We study matrices that share three important features with the dynamical matrix: they are symmetric, positive semidefinite, and force balancing. In higher dimensions, force balance corresponds to $d$ sum rules on partial sums of entries in each row of a matrix, while in 1D, the force balancing restriction simply requires the sum over all the entries in a row must be zero~\citep{Manning2011}. This rule is also obeyed by standard or weighted Laplacians, $L_{ij}$, which are also symmetric and positive semi-definite. They are defined by 
\begin{equation}
L_{ij}=\begin{cases}
-k_{ij}&i\text{ and }j\text{ are connected},\\
\sum_{l\neq i}k_{ik} & i=j,\\
0&\text{Otherwise},
\end{cases}
\end{equation}
where $k_{ij}$ is the independently chosen random weight of the edge between particles $i$ and $j$ and in the special case of the standard Laplacian, $k_{ij}=1$~\citep{Merris1994}. Standard Laplacian matrices are well-studied and possess distinctive vibrational spectra~\citep{Aspelmeier2011}, so we focus on weighted Laplacians for the remainder of this Letter.

In order to calculate the Laplacian we must specify the topology of the underlying graph. Although recent advances have been made in analytically characterizing the spectra of Laplacians on an Erd\H os-R\'enyi graph~\citep{Cicuta2018}, Erd\H os-R\'enyi networks are not locally isostatic, as a significant fraction of nodes are under-coordinated (fewer than isostatic coordination $z_c=2d$), which leads to highly localized excitations that are not seen in jammed packings. 

Instead, we consider the weighted Laplacian on a $z_c$-regular graph with a small number of additional edges, or crossbonds. Since weighted Laplacians only obey one sum rule, they are effectively 1D and $z_c=2$. The number of additional bonds is $\delta zN$ where $N$ is the number of points and $\delta z$ is the excess coordination. 

Another important control parameter is the distribution of the edge weights and, in particular, the weight of this distribution near zero. We choose to parameterize this distribution as a power law with exponent $\alpha$, normalized so that the mean is 1, $\rho(k)\propto k^\alpha$ on $\left[0,\frac{\alpha+2}{\alpha+1}\right]$. A uniform distribution corresponds to $\alpha=0$ and we only consider normalizable distributions, $\alpha>-1$.

{\bf \emph{Finite size scaling for the weighted ring}}: We first study the finite size scaling of the low frequency excitations at isostaticity, when $\delta z=0$ and the underlying network topology is simply a ring of size $N$. Although this is a well-studied model, we believe its finite-size scaling can provide insight into the case with $\delta z>0$.

\begin{figure}[h!]
	\begin{center}
   \includegraphics[width=3.3in]{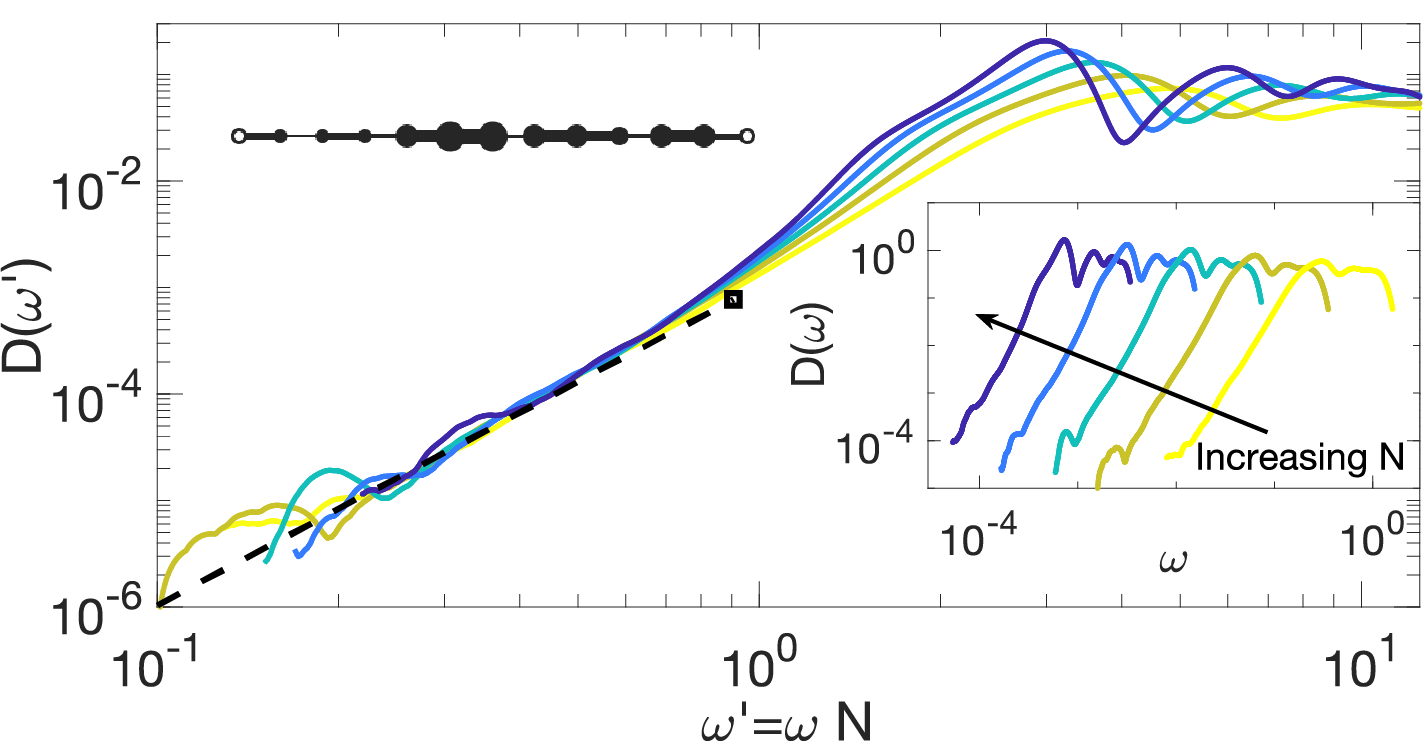}
  \end{center}
  \caption{The rescaled density of states, $D(\omega')$, for the two-regular graph with $N$=16, 64, 256, 1024, and 4096 and $\alpha=0$, normalized by system size, $N$, averaged over at least $10^6$ matrices. The analytic prediction for the low-frequency scaling is shown as the black dashed line. 
In the upper-left we have a sketch of a 1d chain with periodic boundary conditions (the open circles are the same node)
\textbf{Inset:} Unscaled density of states, $D(\omega)$.}
  \label{fig:dos}
\end{figure}

The inset to Fig. \ref{fig:dos} shows the sample averaged density of states for $\alpha=0$, calculated via diagonalization of the matrix, as a function of system size $N$, averaged over $2\times 10^6$ matrices. The main panel shows the sample averaged density of states as a function of the normalized frequency, $\omega'=\omega N$, highlighting a region of power-law scaling at the lowest frequencies that disappears in the thermodynamic limit. 

We hypothesize that the lowest-energy mode on a weighted ring is well approximated by a stretching of the two weakest bonds, with all other bond lengths relatively fixed. We expect this to be the case when $\alpha \le 0$, so that the weight of the lowest two bonds are well separated from bonds with larger values of $k_{ij}$, especially in the limit of low $\omega$, $\omega<N^{-\frac{2\alpha+3}{4 \alpha +3}}$.

If the two weakest bonds have strengths $k_1$ and $k_2$ and are separated by $m$ nodes, the frequency of this mode is $\sqrt{\frac{N(k_1+k_2)}{m(N-m)}}$. As we show in the supplement, one can use extremal statistics to find the exact distribution of the weakest bonds on the ring to predict that the low-frequency density of states scales as:
\begin{equation}
\label{dalpha0}
D(\omega)\propto N^{2\alpha+3}\omega^{4\alpha+3}
\end{equation}

For a uniform distribution of bond weights ($\alpha = 0$), the contribution of these modes to the density of states scales as $(N \omega)^3$. The scaling of Eq. \ref{dalpha0}, using $\alpha =0$, is shown as the black dashed line in Fig. \ref{fig:dos}.

{\bf \emph{Crossbonded ring with uniform bond weights:}} We hypothesize that adding a small number of crossbonds alters the low-frequency behavior by reducing the effective distance between the two weakest bonds. In the case of $\delta z =0$, the two weakest bonds separate the ring into two segments that can move relative to one another at nearly zero cost, but if a crossbond connects those two segments it will significantly increase the energy of that mode. Therefore, the weak bonds that contribute to low-frequency modes must both be in a segment between crossbonds. Because there are $N\delta z$ such segments, we expect that crossbonds give rise to an extensive number of low-energy modes, so that the scaling regime described in the previous section persists in the thermodynamic limit.

We search for very low-weight edges that generate a two-cut of the network: two edges that, if removed, disconnect the network. In the supplement, we show the low-frequency density of states scales as

\begin{equation}
\label{Dalpha}
D_\alpha(\omega)\propto\frac{\omega^{4\alpha+3}}{\delta z^{2\alpha+3}},
\end{equation}
independent of system size.

\begin{figure}[h!]
	\begin{center}
   \includegraphics[width=3.4in]{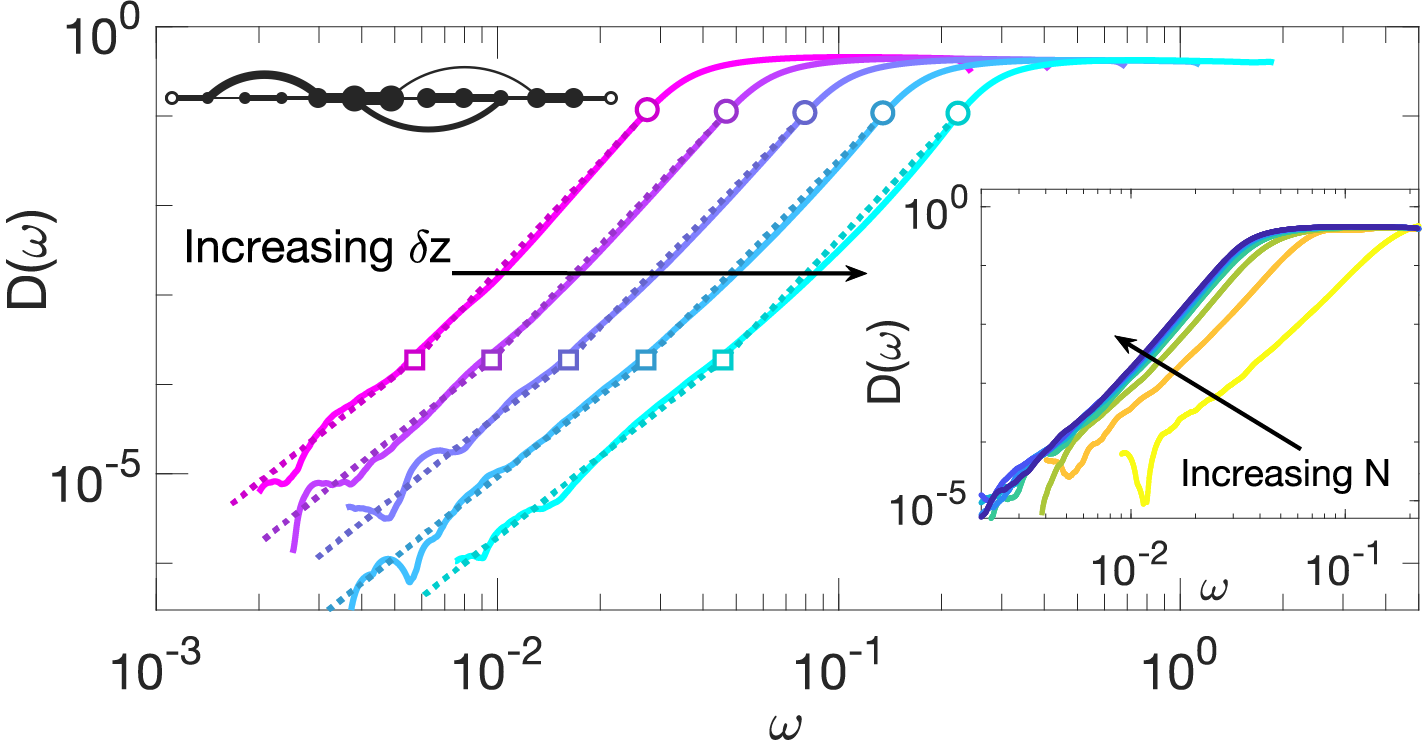}
  \end{center}
  \caption{
The density of states for fixed system size (N=1000) and changing $\delta z=0.1,0.168,0.282,0.476,0.8$ 
In the upper-left we have a sketch of a 1d chain with periodic boundary conditions (the open circles are the same node) with additional bonds.
\textbf{Inset:} The density of states, $D(\omega)$, for fixed $\delta z=0.1$ and changing system size $N=20$, $60$, $120$, $240$, $500$, $1000$, $2000$, and $4000$.
}
  \label{fig:full}
\end{figure}

To test the universal form predicted by Eq. \ref{Dalpha}, we computed the spectrum $D(\omega)$ for rings with crossbonds and uniform bond weights ($\alpha=0$). For each value of $\delta z$ and $N$ we generated between $10^5$ and $2 \times 10^6$ matrices samples~\footnote{ For $\delta z=0.1$ and $N=500$ and $1000$, we calculate $2\times 10^6$ matrices and for $N=2000$ and $4000$, we calculate $522240$ and $261120$ matrices. For all other values, we calculate $10^6$ matrices.}, with independently chosen weights and uniformly random placements of the endpoints of the $N\delta z/2$ crossbonds. The inset to Fig. \ref{fig:full} displays plots of the sample-averaged density of states $D(\omega)$ for fixed $\delta z =0.1$ as $N$ increases. This example plot supports the convergence of $D(\omega)$ to a gapless distribution as $N\rightarrow \infty$. The main panel of Fig. \ref{fig:full} displays the computed density of states (solid lines) for large $N$ ($N=1000$) and varying $\delta z$. The dashed lines in Fig. \ref{fig:full} show fits of the form $D(\omega)\propto\omega^3$ to the low frequency region, as predicted by Eq. \ref{Dalpha}. These fits are in good agreement with the computed spectra.

\begin{figure}[h!]
	\begin{center}
   \includegraphics[width=3.4in]{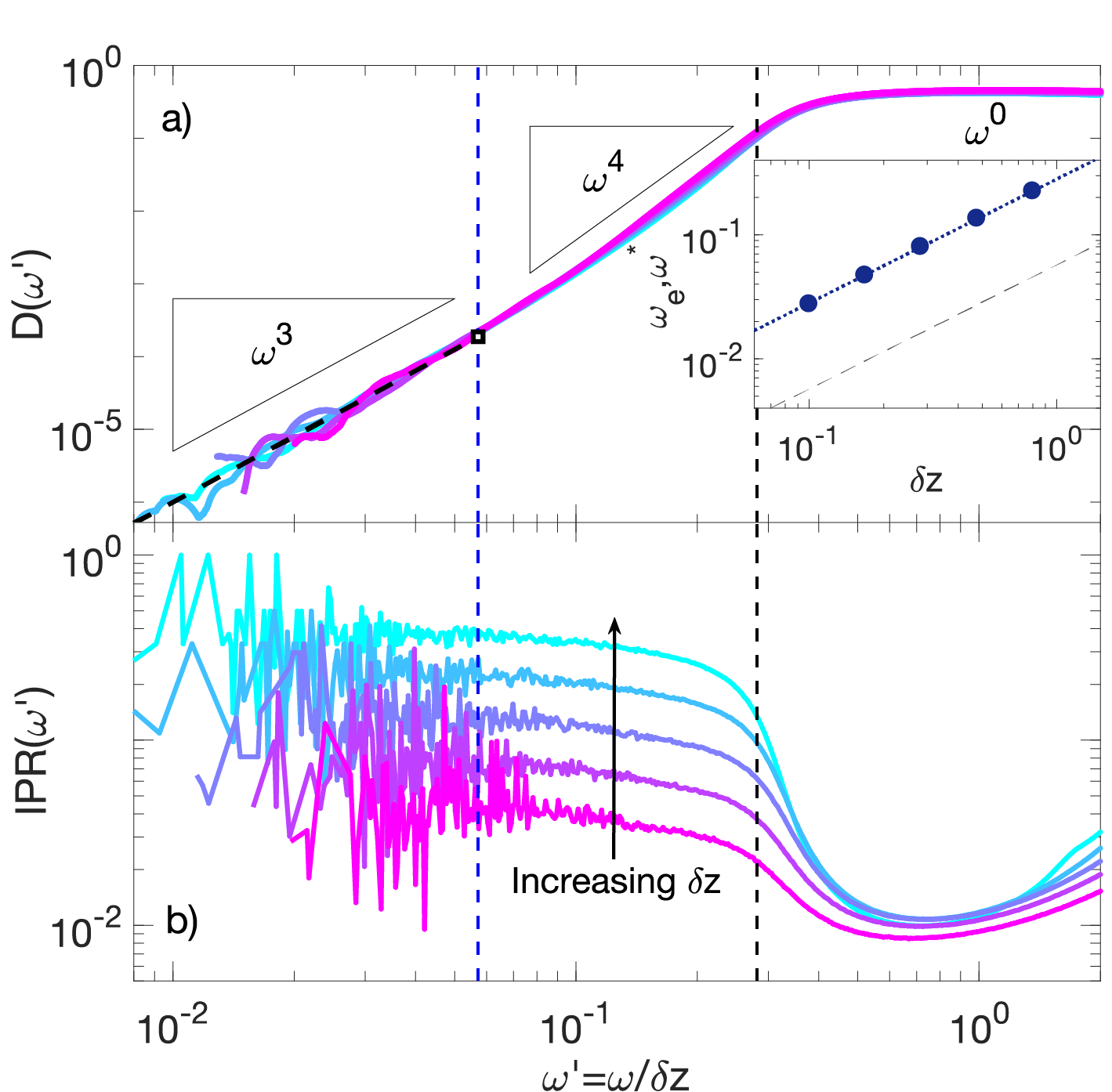}
  \end{center}
  \caption{\textbf{a)}The density of states, $D(\omega)$, rescaled by $\delta z$. The blue dashed line indicates the transition from the $\omega^3$ regime to the $\omega^4$ regime while the black dashed line indicates the transition to the plateau. The inset shows the scaling of $\omega^*$ and $\omega_e$ with $\delta z$ is linear.
\textbf{b)}The inverse participation ratio, IPR, rescaled by $\delta z$. The IPR approaches a quasilocalized plateau in the $\omega^3$ region.
}
  \label{fig:full2}
\end{figure}

Based on Eq. \ref{Dalpha} and the more complete form of the density of states derived in Appendix B, we expect a collapse of $D(\omega)$ when frequencies are scaled by $\delta z$. Fig. \ref{fig:full2}(a) shows the density of states for the scaled frequency, $\omega’=\omega/\delta z$. For $\delta z = 0.168$ we numerically identify a frequency $\omega_e$ that best separates the $\omega^3$ scaling regime from the remaining spectrum. Eq.~\ref{Dalpha} then predicts that all other cutoff frequencies should scale linearly with $\delta z$, which is in good agreement with the data as shown by the open squares in Fig~\ref{fig:full} and~\ref{fig:full2}(a).

In addition to the crossover at $\omega_e$, there is a second crossover where $D(\omega)$ flattens to a plateau. In jammed packings at zero temperature, where the boson peak occurs at the onset of the plateau, $\omega^*$ is often defined as the frequency at which the density of states attains a fixed fraction $f$ (typically 25 \%) of its value in the plateau~~\citep{Xu2010}. We use that same definition here with $f=0.25$.  

In many disordered solids, numerical evidence suggests $\omega^*\propto\delta z$~\citep{OhernAPS,Silbert2005}. To check whether this is also true for our matrices, we plot the density of states as a function of the rescaled frequency $\omega' = \omega/ \delta z$, for various values of $\delta z$, shown in Fig~\ref{fig:full2}(a). We see a good collapse of the three regions, suggesting that both crossovers are linear in $\delta z$, which is also highlighted by the inset to Fig~\ref{fig:full2}(a). 

Importantly, this confirms that although the intermediate region between the two crossover frequencies spans less than a decade in frequency, it is well-defined and does not change as a function of excess coordination or system size. Specifically, these results mandate the following functional form for the density of states in our random matrix model with $\alpha = 0$:

\begin{equation}
\label{scaling}
D(\omega)=\begin{cases}
\frac{4}{L^2} \left( \frac{\omega}{\delta z}\right)^3 &\omega\leq \omega_e\\
\propto \omega^\psi & \omega_e\leq\omega\leq\omega^*\\
\propto \omega^0 & \omega^*\leq\omega
\end{cases}
\end{equation}

To extract the scaling of $D(\omega)$ below the boson peak, we fit $D(\omega)$ to this functional form and extract the best-fit $\psi$ for each value of $\delta z$ (See table in supplemental materials). We find that all curves are consistent with $\psi = 4.0\pm 0.05$ for frequencies $\omega_e\le \omega \le \omega^*$. This suggests $D(\omega) \propto \omega^4$, just as seen below the plateau in simulations of jammed packings.

Given the striking similarities between the density of states in this simple model and jammed packings, we would also like to know if the eigenvector statistics are similar. In jammed systems, many modes at frequencies below the boson peak are quasilocalized~\citep{Xu2010}. This is quantified by the inverse participation ratio (IPR), $IPR(\omega)= \sum_i v_i^4 / (\sum_i v_i^2)^2$, where $v$ is the vector associated with the eigenfrequency $\omega$. In Fig \ref{fig:full}(b), the very low-frequency regime of the IPR plateaus, and the value of this plateau scales with $\delta z$, indicating that only about $\frac{1}{\delta z}$ nodes are participating in the vibration. 

Interestingly, the intermediate region exhibits values of IPR that are typically associated with quasilocalized excitations. Moreover, the size of those excitations seems to decrease as $\delta z$ increases. In jammed solids, an outstanding open question is how the size of localized excitations changes as one approaches the jamming transition.

\emph{\bf \emph{Crossbonded ring with power-law bond weights}}: Having a simple constructive model that reproduces many features of the vibrational modes in jammed packings is useful, because we can vary the model and ask what features are necessary to generate the $\omega^4$ scaling in the density of states. One natural choice is to perturb the distribution of bond strengths away from the uniform distribution by changing the power-law exponent $\alpha$. 

\begin{figure}[h!]
	\begin{center}
   \includegraphics[width=3.3in]{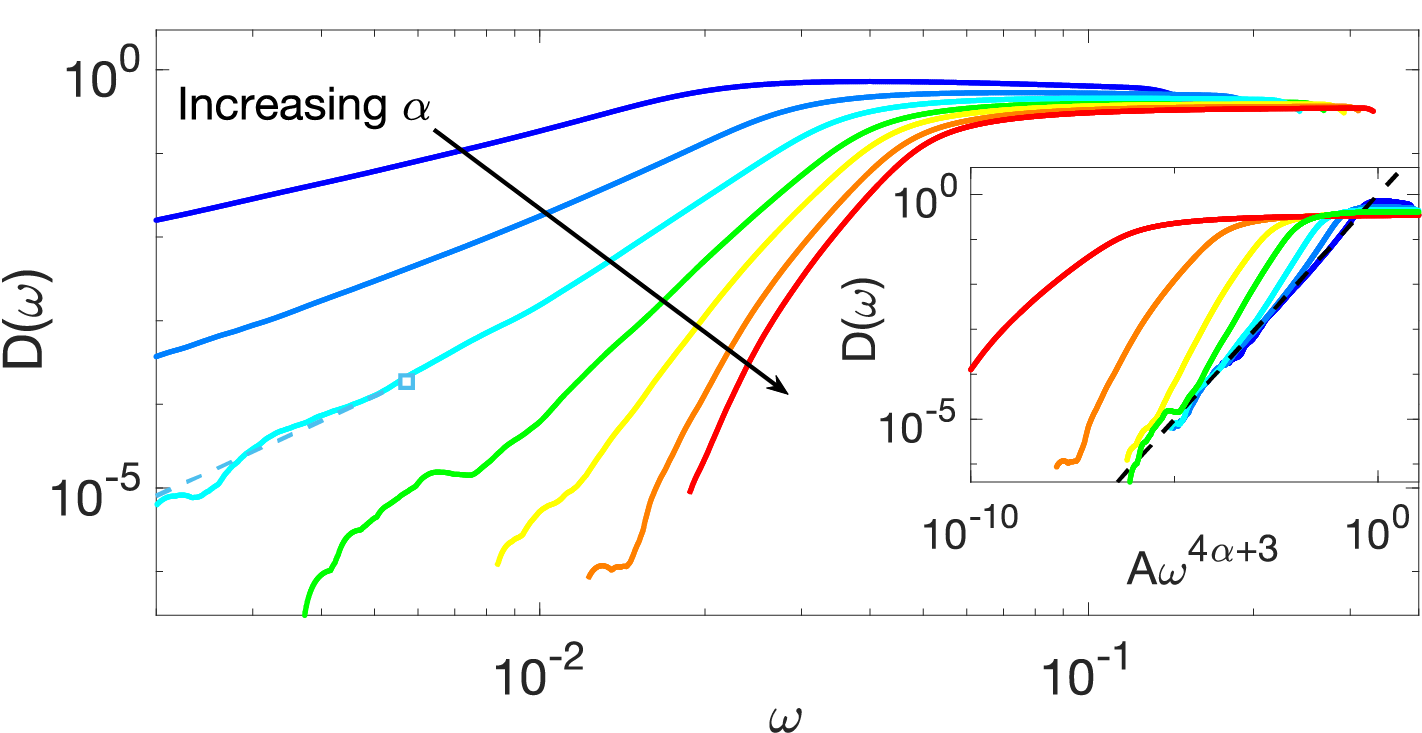}
  \end{center}
  \caption{The density of states for $\alpha=-0.4$, $-0.2$, $0$, $0.25$, $0.5$, $1$, and $2$, with $\delta z=0.1$.
\textbf{Inset:} $D(\omega'=A \omega^{4\alpha+3})$ for the same values of $\alpha$ as in the main figure, where $A$ is the coefficient predicted in Appendix B. The black dashed line is the predicted scaling for the low frequency regime.
}
  \label{fig:alpha}
\end{figure}

For $\alpha>0$, very weak bonds become rare and the assumptions that lead to Eq.~\ref{Dalpha} break down. Numerically, we observe that a gap appears to open up in the spectrum as $\alpha$ increases, as seen in Fig \ref{fig:alpha}. For $\alpha<0$, we expect Eq.~\ref{Dalpha} should still hold, as shown by the numerical data in the inset of Fig \ref{fig:alpha}. In this case, however, the crossover frequency no longer scales linearly with $\delta z$, and so the power-law scaling between $\omega_e$ and $\omega^*$ -- the exponent $\psi$ in Eq.~\ref{scaling} -- is no longer independent of $\delta z$. In other words, an intermediate regime consistent with $D(\omega) \propto \omega^4$, independent of $\delta z$, is only possible for $\alpha=0$.

{\bf Discussion:} In this Letter, we propose a simple random matrix model that is locally nearly isostatic and captures features of the vibrational states of disordered packings that are typically associated with marginality. Specifically, the model recapitulates a plateau in the density of states above $\omega^*$, and a regime consistent with $\omega^4$ scaling immediately below that. Our model also has a second crossover frequency $\omega_e$, below which $D(\omega)$ scales as $\omega^3$.

The modes in this extremely low frequency regime are governed by extremal statistics, and so we can calculate their properties analytically. This allows us to demonstrate that $\omega_e$ scales linearly with excess coordination $\delta z$ if and only if the weak bonds are uniformly distributed, suggesting that $\omega^4$ seen in jammed packings arises due to a special, self-organized distribution of the weakest bonds.

Of course, jammed packings only exist in dimensions greater than unity. Above one dimension, the bond between particles is described by a tensor and not a scalar weight. The $d$ by $d$ interaction block that corresponds to a single bond in the Hessian matrix can be written as $H_{ij\alpha\beta}=-V''|u_\parallel|^2-\frac{V'}{r_{ij}}|u_\perp|^2$. The first term is often referred to as the stiffness while the second term is called the prestress term~\citep{Ellenbroek}.

Interestingly, observations in 3D jammed packings suggest that the $\omega^4$ regime only exists when the $V'$ term is unperturbed; even very small perturbations to the prestress open up a gap in the density of states~\citep{Lerner2017}. This suggests that a self-organized balance between the stiffness and prestress must occur in systems near isostaticity. Moreover, the stiffness is always positive and the prestress always decreases the entries in the Hessian, so it is plausible that the prestress term is driving some interactions to be very weak near isostaticity, similar to our simple model. 

While suggestive, a more concrete connection will require us to extend our analysis to higher dimensions. We see an $\omega^4$ regime when bond strengths are uniform, but it is unclear what quantity would be analogous to a uniform bond weight in a $d\times d$ sub-block. Concurrent work by Benetti et al focused on $d$-dimensional Laplacian matrices where the magnitude of each bond is unity, but the geometry of the bond is randomly distributed, and these also generate scaling consistent with $\omega^4$ at low frequencies~\citep{Benetti2018}. To better understand the connections between these models and why both generate $\omega^4$ scaling, one could study systems with random bond weights and ordered geometries, or have both be disordered. 

Furthermore, although $\omega^4$ scaling as been observed in several glass forming systems~\citep{Lerner2016}, the $\omega^3$ regime may be unique to 1D systems, as it has not been reported in simulations or in the random matrices with $3 \times 3$ sub-blocks~\cite{Benetti2018}. In addition, we see about half a decade of frequency consistent with $\omega^4$ scaling, while the most recent data from Lerner and collaborators~\citep{Lerner2016,Lerner2017,Kapteijns2018} finds almost a full decade. 

Nevertheless, the $\omega^3$ scaling regime is interesting. Disordered rings are well-studied, but major results focus on localization caused by disorder~\citep{Dyson1953,Dean1964}. To our knowledge, the finite-size scaling effects of the vibrational spectrum have not been discussed previously. Our model demonstrates that finite size effects in the disordered ring, such as this gapless low-frequency scaling, can be promoted into properties that are maintained in the thermodynamic limit by network disorder. 

Although we have excellent understanding of the $\omega^3$ regime in this simple model, and convincing numerical evidence demonstrating $D(\omega)$ scaling as $\omega^4$ over a window of about half of a decade in $\omega$, we have not identified a mechanism for the $\omega^4$ regime, where we know the assumption of two weak bonds and two rigid arms breaks down. There are many higher order modes that may contribute, 
and visual inspection of the eigenvector structure suggests that no single one dominates, so there is no obvious simple extension of our argument for $\omega^3$.

One possible avenue for understanding this regime is suggested by recent numerical work that shows universality in the eigenvector statistics associated with the boson peak. Specifically, eigenstatistics in jammed packings match those from both the random matrix model described here, as well as the dense limit of this model where all nodes are connected to one another~\citep{Manning2015}. Interestingly, the eigenvector statistics are also identical in a much simpler model which is just the sum of a diagonal matrix and a Gaussian orthogonal matrix. Very recent analytic work suggests that such matrices are marginal; they are on the edge of a non-ergodic localized phase~\citep{Facoetti2016a}. It would therefore be very interesting to extend this analytic work to sparse matrices and study the tail of the density of states.

Another way to extend our model is to alter the loop structure of the underlying graph. In our random matrix model, the loop structure is uncontrolled since we add crossbonds with uniform probability across the graph. This is different from jammed systems where neighbors of one particle are more likely to be neighbors of each other and loops are small. It is fairly straightforward to extend our analytic analysis of the $\omega^3$ regime to random matrix models with smaller loops, and we expect that the prefactor and the onset of the scaling $\omega_e$ will change, but the $\omega^3$ scaling will not. However, this change could impact the behavior of the $\omega^4$ regime.

{\bf Acknowledgements} We thank Fernanda Benetti, Gabriele Sicuro, and Giorgio Parisi for discussions. 
This work was partially supported by the Simons Foundation grant number 454947 (ES, PM, MLM), and by NSF-DMR-1352184 (ES, MLM). Computational resources were provided by support from Syracuse University and NSF ACI-1541396.

\bibliography{RM.bib}

\begin{thebibliography}{31}%
\makeatletter
\providecommand \@ifxundefined [1]{%
 \@ifx{#1\undefined}
}%
\providecommand \@ifnum [1]{%
 \ifnum #1\expandafter \@firstoftwo
 \else \expandafter \@secondoftwo
 \fi
}%
\providecommand \@ifx [1]{%
 \ifx #1\expandafter \@firstoftwo
 \else \expandafter \@secondoftwo
 \fi
}%
\providecommand \natexlab [1]{#1}%
\providecommand \enquote  [1]{``#1''}%
\providecommand \bibnamefont  [1]{#1}%
\providecommand \bibfnamefont [1]{#1}%
\providecommand \citenamefont [1]{#1}%
\providecommand \href@noop [0]{\@secondoftwo}%
\providecommand \href [0]{\begingroup \@sanitize@url \@href}%
\providecommand \@href[1]{\@@startlink{#1}\@@href}%
\providecommand \@@href[1]{\endgroup#1\@@endlink}%
\providecommand \@sanitize@url [0]{\catcode `\\12\catcode `\$12\catcode
  `\&12\catcode `\#12\catcode `\^12\catcode `\_12\catcode `\%12\relax}%
\providecommand \@@startlink[1]{}%
\providecommand \@@endlink[0]{}%
\providecommand \url  [0]{\begingroup\@sanitize@url \@url }%
\providecommand \@url [1]{\endgroup\@href {#1}{\urlprefix }}%
\providecommand \urlprefix  [0]{URL }%
\providecommand \Eprint [0]{\href }%
\providecommand \doibase [0]{http://dx.doi.org/}%
\providecommand \selectlanguage [0]{\@gobble}%
\providecommand \bibinfo  [0]{\@secondoftwo}%
\providecommand \bibfield  [0]{\@secondoftwo}%
\providecommand \translation [1]{[#1]}%
\providecommand \BibitemOpen [0]{}%
\providecommand \bibitemStop [0]{}%
\providecommand \bibitemNoStop [0]{.\EOS\space}%
\providecommand \EOS [0]{\spacefactor3000\relax}%
\providecommand \BibitemShut  [1]{\csname bibitem#1\endcsname}%
\let\auto@bib@innerbib\@empty
\bibitem [{\citenamefont {Manning}\ and\ \citenamefont
  {Liu}(2011)}]{Manning2011}%
  \BibitemOpen
  \bibfield  {author} {\bibinfo {author} {\bibfnamefont {M.~L.}\ \bibnamefont
  {Manning}}\ and\ \bibinfo {author} {\bibfnamefont {A.~J.}\ \bibnamefont
  {Liu}},\ }\href {https://doi.org/10.1103/physrevlett.107.108302} {\bibfield
  {journal} {\bibinfo  {journal} {Physical Review Letters}\ }\textbf {\bibinfo
  {volume} {107}} (\bibinfo {year} {2011})}\BibitemShut {NoStop}%
\bibitem [{\citenamefont {O'Hern}\ \emph {et~al.}(2003)\citenamefont {O'Hern},
  \citenamefont {Silbert}, \citenamefont {Liu},\ and\ \citenamefont
  {Nagel}}]{OhernAPS}%
  \BibitemOpen
  \bibfield  {author} {\bibinfo {author} {\bibfnamefont {C.~S.}\ \bibnamefont
  {O'Hern}}, \bibinfo {author} {\bibfnamefont {L.~E.}\ \bibnamefont {Silbert}},
  \bibinfo {author} {\bibfnamefont {A.~J.}\ \bibnamefont {Liu}}, \ and\
  \bibinfo {author} {\bibfnamefont {S.~R.}\ \bibnamefont {Nagel}},\ }\href
  {https://doi.org/10.1103/physreve.68.011306} {\bibfield  {journal} {\bibinfo
  {journal} {Physical Review E}\ }\textbf {\bibinfo {volume} {68}} (\bibinfo
  {year} {2003})}\BibitemShut {NoStop}%
\bibitem [{\citenamefont {DeGiuli}\ \emph {et~al.}(2014)\citenamefont
  {DeGiuli}, \citenamefont {Laversanne-Finot}, \citenamefont {D\"{u}ring},
  \citenamefont {Lerner},\ and\ \citenamefont {Wyart}}]{DeGiuli2014}%
  \BibitemOpen
  \bibfield  {author} {\bibinfo {author} {\bibfnamefont {E.}~\bibnamefont
  {DeGiuli}}, \bibinfo {author} {\bibfnamefont {A.}~\bibnamefont
  {Laversanne-Finot}}, \bibinfo {author} {\bibfnamefont {G.}~\bibnamefont
  {D\"{u}ring}}, \bibinfo {author} {\bibfnamefont {E.}~\bibnamefont {Lerner}},
  \ and\ \bibinfo {author} {\bibfnamefont {M.}~\bibnamefont {Wyart}},\ }\href
  {https://doi.org/10.1039/c4sm00561a} {\bibfield  {journal} {\bibinfo
  {journal} {Soft Matter}\ }\textbf {\bibinfo {volume} {10}},\ \bibinfo {pages}
  {5628} (\bibinfo {year} {2014})}\BibitemShut {NoStop}%
\bibitem [{\citenamefont {Wyart}\ \emph {et~al.}(2005)\citenamefont {Wyart},
  \citenamefont {Nagel},\ and\ \citenamefont {Witten}}]{Wyart2005}%
  \BibitemOpen
  \bibfield  {author} {\bibinfo {author} {\bibfnamefont {M.}~\bibnamefont
  {Wyart}}, \bibinfo {author} {\bibfnamefont {S.~R.}\ \bibnamefont {Nagel}}, \
  and\ \bibinfo {author} {\bibfnamefont {T.~A.}\ \bibnamefont {Witten}},\
  }\href {https://doi.org/10.1209/epl/i2005-10245-5} {\bibfield  {journal}
  {\bibinfo  {journal} {Europhysics Letters ({EPL})}\ }\textbf {\bibinfo
  {volume} {72}},\ \bibinfo {pages} {486} (\bibinfo {year} {2005})}\BibitemShut
  {NoStop}%
\bibitem [{\citenamefont {Silbert}\ \emph {et~al.}(2005)\citenamefont
  {Silbert}, \citenamefont {Liu},\ and\ \citenamefont {Nagel}}]{Silbert2005}%
  \BibitemOpen
  \bibfield  {author} {\bibinfo {author} {\bibfnamefont {L.~E.}\ \bibnamefont
  {Silbert}}, \bibinfo {author} {\bibfnamefont {A.~J.}\ \bibnamefont {Liu}}, \
  and\ \bibinfo {author} {\bibfnamefont {S.~R.}\ \bibnamefont {Nagel}},\ }\href
  {https://doi.org/10.1103/physrevlett.95.098301} {\bibfield  {journal}
  {\bibinfo  {journal} {Physical Review Letters}\ }\textbf {\bibinfo {volume}
  {95}} (\bibinfo {year} {2005})}\BibitemShut {NoStop}%
\bibitem [{\citenamefont {Manning}\ and\ \citenamefont
  {Liu}(2015)}]{Manning2015}%
  \BibitemOpen
  \bibfield  {author} {\bibinfo {author} {\bibfnamefont {M.~L.}\ \bibnamefont
  {Manning}}\ and\ \bibinfo {author} {\bibfnamefont {A.~J.}\ \bibnamefont
  {Liu}},\ }\href {https://doi.org/10.1209/0295-5075/109/36002} {\bibfield
  {journal} {\bibinfo  {journal} {{EPL} (Europhysics Letters)}\ }\textbf
  {\bibinfo {volume} {109}},\ \bibinfo {pages} {36002} (\bibinfo {year}
  {2015})}\BibitemShut {NoStop}%
\bibitem [{\citenamefont {Lerner}\ \emph {et~al.}(2016)\citenamefont {Lerner},
  \citenamefont {D\"{u}ring},\ and\ \citenamefont {Bouchbinder}}]{Lerner2016}%
  \BibitemOpen
  \bibfield  {author} {\bibinfo {author} {\bibfnamefont {E.}~\bibnamefont
  {Lerner}}, \bibinfo {author} {\bibfnamefont {G.}~\bibnamefont {D\"{u}ring}},
  \ and\ \bibinfo {author} {\bibfnamefont {E.}~\bibnamefont {Bouchbinder}},\
  }\href {https://doi.org/10.1103/physrevlett.117.035501} {\bibfield  {journal}
  {\bibinfo  {journal} {Physical Review Letters}\ }\textbf {\bibinfo {volume}
  {117}} (\bibinfo {year} {2016})}\BibitemShut {NoStop}%
\bibitem [{\citenamefont {Mizuno}\ \emph {et~al.}(2017)\citenamefont {Mizuno},
  \citenamefont {Shiba},\ and\ \citenamefont {Ikeda}}]{Mizuno2017}%
  \BibitemOpen
  \bibfield  {author} {\bibinfo {author} {\bibfnamefont {H.}~\bibnamefont
  {Mizuno}}, \bibinfo {author} {\bibfnamefont {H.}~\bibnamefont {Shiba}}, \
  and\ \bibinfo {author} {\bibfnamefont {A.}~\bibnamefont {Ikeda}},\ }\href
  {https://doi.org/10.1073/pnas.1709015114} {\bibfield  {journal} {\bibinfo
  {journal} {Proceedings of the National Academy of Sciences}\ }\textbf
  {\bibinfo {volume} {114}},\ \bibinfo {pages} {E9767} (\bibinfo {year}
  {2017})}\BibitemShut {NoStop}%
\bibitem [{\citenamefont {Kapteijns}\ \emph {et~al.}(2018)\citenamefont
  {Kapteijns}, \citenamefont {Bouchbinder},\ and\ \citenamefont
  {Lerner}}]{Kapteijns2018}%
  \BibitemOpen
  \bibfield  {author} {\bibinfo {author} {\bibfnamefont {G.}~\bibnamefont
  {Kapteijns}}, \bibinfo {author} {\bibfnamefont {E.}~\bibnamefont
  {Bouchbinder}}, \ and\ \bibinfo {author} {\bibfnamefont {E.}~\bibnamefont
  {Lerner}},\ }\href@noop {} {\enquote {\bibinfo {title} {Universal
  non-phononic density of states in 2d, 3d and 4d glasses},}\ } (\bibinfo
  {year} {2018}),\ \Eprint {http://arxiv.org/abs/arXiv:1803.11383}
  {arXiv:1803.11383} \BibitemShut {NoStop}%
\bibitem [{\citenamefont {Charbonneau}\ \emph {et~al.}(2016)\citenamefont
  {Charbonneau}, \citenamefont {Corwin}, \citenamefont {Parisi}, \citenamefont
  {Poncet},\ and\ \citenamefont {Zamponi}}]{Charbonneau2016}%
  \BibitemOpen
  \bibfield  {author} {\bibinfo {author} {\bibfnamefont {P.}~\bibnamefont
  {Charbonneau}}, \bibinfo {author} {\bibfnamefont {E.~I.}\ \bibnamefont
  {Corwin}}, \bibinfo {author} {\bibfnamefont {G.}~\bibnamefont {Parisi}},
  \bibinfo {author} {\bibfnamefont {A.}~\bibnamefont {Poncet}}, \ and\ \bibinfo
  {author} {\bibfnamefont {F.}~\bibnamefont {Zamponi}},\ }\href
  {https://doi.org/10.1103/physrevlett.117.045503} {\bibfield  {journal}
  {\bibinfo  {journal} {Physical Review Letters}\ }\textbf {\bibinfo {volume}
  {117}} (\bibinfo {year} {2016})}\BibitemShut {NoStop}%
\bibitem [{\citenamefont {Parisi}\ and\ \citenamefont
  {Zamponi}(2010)}]{Parisi2010}%
  \BibitemOpen
  \bibfield  {author} {\bibinfo {author} {\bibfnamefont {G.}~\bibnamefont
  {Parisi}}\ and\ \bibinfo {author} {\bibfnamefont {F.}~\bibnamefont
  {Zamponi}},\ }\href {https://doi.org/10.1103/revmodphys.82.789} {\bibfield
  {journal} {\bibinfo  {journal} {Reviews of Modern Physics}\ }\textbf
  {\bibinfo {volume} {82}},\ \bibinfo {pages} {789} (\bibinfo {year}
  {2010})}\BibitemShut {NoStop}%
\bibitem [{\citenamefont {Baity-Jesi}\ \emph {et~al.}(2015)\citenamefont
  {Baity-Jesi}, \citenamefont {Mart{\'{\i}}n-Mayor}, \citenamefont {Parisi},\
  and\ \citenamefont {Perez-Gaviro}}]{BaityJesi2015}%
  \BibitemOpen
  \bibfield  {author} {\bibinfo {author} {\bibfnamefont {M.}~\bibnamefont
  {Baity-Jesi}}, \bibinfo {author} {\bibfnamefont {V.}~\bibnamefont
  {Mart{\'{\i}}n-Mayor}}, \bibinfo {author} {\bibfnamefont {G.}~\bibnamefont
  {Parisi}}, \ and\ \bibinfo {author} {\bibfnamefont {S.}~\bibnamefont
  {Perez-Gaviro}},\ }\href {\doibase 10.1103/physrevlett.115.267205} {\bibfield
   {journal} {\bibinfo  {journal} {Physical Review Letters}\ }\textbf {\bibinfo
  {volume} {115}} (\bibinfo {year} {2015}),\
  10.1103/physrevlett.115.267205}\BibitemShut {NoStop}%
\bibitem [{\citenamefont {Wijtmans}\ and\ \citenamefont
  {Manning}(2017)}]{Wijtmans2017}%
  \BibitemOpen
  \bibfield  {author} {\bibinfo {author} {\bibfnamefont {S.}~\bibnamefont
  {Wijtmans}}\ and\ \bibinfo {author} {\bibfnamefont {M.~L.}\ \bibnamefont
  {Manning}},\ }\href {https://doi.org/10.1039/c7sm00792b} {\bibfield
  {journal} {\bibinfo  {journal} {Soft Matter}\ }\textbf {\bibinfo {volume}
  {13}},\ \bibinfo {pages} {5649} (\bibinfo {year} {2017})}\BibitemShut
  {NoStop}%
\bibitem [{\citenamefont {Tanguy}\ \emph {et~al.}(2010)\citenamefont {Tanguy},
  \citenamefont {Mantisi},\ and\ \citenamefont {Tsamados}}]{Tanguy2010}%
  \BibitemOpen
  \bibfield  {author} {\bibinfo {author} {\bibfnamefont {A.}~\bibnamefont
  {Tanguy}}, \bibinfo {author} {\bibfnamefont {B.}~\bibnamefont {Mantisi}}, \
  and\ \bibinfo {author} {\bibfnamefont {M.}~\bibnamefont {Tsamados}},\ }\href
  {https://doi.org/10.1209/0295-5075/90/16004} {\bibfield  {journal} {\bibinfo
  {journal} {{EPL} (Europhysics Letters)}\ }\textbf {\bibinfo {volume} {90}},\
  \bibinfo {pages} {16004} (\bibinfo {year} {2010})}\BibitemShut {NoStop}%
\bibitem [{\citenamefont {Tsamados}\ \emph {et~al.}(2008)\citenamefont
  {Tsamados}, \citenamefont {Tanguy}, \citenamefont {L{\'{e}}onforte},\ and\
  \citenamefont {Barrat}}]{Tsamados2008}%
  \BibitemOpen
  \bibfield  {author} {\bibinfo {author} {\bibfnamefont {M.}~\bibnamefont
  {Tsamados}}, \bibinfo {author} {\bibfnamefont {A.}~\bibnamefont {Tanguy}},
  \bibinfo {author} {\bibfnamefont {F.}~\bibnamefont {L{\'{e}}onforte}}, \ and\
  \bibinfo {author} {\bibfnamefont {J.~L.}\ \bibnamefont {Barrat}},\ }\href
  {https://doi.org/10.1140/epje/i2007-10324-y} {\bibfield  {journal} {\bibinfo
  {journal} {The European Physical Journal E}\ }\textbf {\bibinfo {volume}
  {26}},\ \bibinfo {pages} {283} (\bibinfo {year} {2008})}\BibitemShut
  {NoStop}%
\bibitem [{\citenamefont {Ashton}\ and\ \citenamefont
  {Garrahan}(2009)}]{Ashton2009}%
  \BibitemOpen
  \bibfield  {author} {\bibinfo {author} {\bibfnamefont {D.~J.}\ \bibnamefont
  {Ashton}}\ and\ \bibinfo {author} {\bibfnamefont {J.~P.}\ \bibnamefont
  {Garrahan}},\ }\href {\doibase 10.1140/epje/i2009-10531-6} {\bibfield
  {journal} {\bibinfo  {journal} {The European Physical Journal E}\ }\textbf
  {\bibinfo {volume} {30}} (\bibinfo {year} {2009}),\
  10.1140/epje/i2009-10531-6}\BibitemShut {NoStop}%
\bibitem [{\citenamefont {Brito}\ and\ \citenamefont
  {Wyart}(2007)}]{Brito2007}%
  \BibitemOpen
  \bibfield  {author} {\bibinfo {author} {\bibfnamefont {C.}~\bibnamefont
  {Brito}}\ and\ \bibinfo {author} {\bibfnamefont {M.}~\bibnamefont {Wyart}},\
  }\href {https://doi.org/10.1088/1742-5468/2007/08/l08003} {\bibfield
  {journal} {\bibinfo  {journal} {Journal of Statistical Mechanics: Theory and
  Experiment}\ }\textbf {\bibinfo {volume} {2007}},\ \bibinfo {pages} {L08003}
  (\bibinfo {year} {2007})}\BibitemShut {NoStop}%
\bibitem [{\citenamefont {Mehta}(2004)}]{mehta2004}%
  \BibitemOpen
  \bibfield  {author} {\bibinfo {author} {\bibfnamefont {M.}~\bibnamefont
  {Mehta}},\ }\href {https://books.google.com/books?id=Kp3Nx03\_gMwC} {\emph
  {\bibinfo {title} {Random Matrices}}},\ Pure and Applied Mathematics\
  (\bibinfo  {publisher} {Elsevier Science},\ \bibinfo {year}
  {2004})\BibitemShut {NoStop}%
\bibitem [{\citenamefont {Parisi}(2002)}]{Parisi2002}%
  \BibitemOpen
  \bibfield  {author} {\bibinfo {author} {\bibfnamefont {G.}~\bibnamefont
  {Parisi}},\ }\href {https://doi.org/10.1140/epje/i2002-10088-x} {\bibfield
  {journal} {\bibinfo  {journal} {The European Physical Journal E - Soft
  Matter}\ }\textbf {\bibinfo {volume} {9}},\ \bibinfo {pages} {213} (\bibinfo
  {year} {2002})}\BibitemShut {NoStop}%
\bibitem [{\citenamefont {Gurarie}\ and\ \citenamefont
  {Chalker}(2003)}]{Gurarie2003}%
  \BibitemOpen
  \bibfield  {author} {\bibinfo {author} {\bibfnamefont {V.}~\bibnamefont
  {Gurarie}}\ and\ \bibinfo {author} {\bibfnamefont {J.~T.}\ \bibnamefont
  {Chalker}},\ }\href {https://doi.org/10.1103/physrevb.68.134207} {\bibfield
  {journal} {\bibinfo  {journal} {Physical Review B}\ }\textbf {\bibinfo
  {volume} {68}} (\bibinfo {year} {2003})}\BibitemShut {NoStop}%
\bibitem [{\citenamefont {Merris}(1994)}]{Merris1994}%
  \BibitemOpen
  \bibfield  {author} {\bibinfo {author} {\bibfnamefont {R.}~\bibnamefont
  {Merris}},\ }\href {https://doi.org/10.1016/0024-3795(94)90486-3} {\bibfield
  {journal} {\bibinfo  {journal} {Linear Algebra and its Applications}\
  }\textbf {\bibinfo {volume} {197-198}},\ \bibinfo {pages} {143} (\bibinfo
  {year} {1994})}\BibitemShut {NoStop}%
\bibitem [{\citenamefont {Aspelmeier}\ and\ \citenamefont
  {Zippelius}(2011)}]{Aspelmeier2011}%
  \BibitemOpen
  \bibfield  {author} {\bibinfo {author} {\bibfnamefont {T.}~\bibnamefont
  {Aspelmeier}}\ and\ \bibinfo {author} {\bibfnamefont {A.}~\bibnamefont
  {Zippelius}},\ }\href {https://doi.org/10.1007/s10955-011-0271-2} {\bibfield
  {journal} {\bibinfo  {journal} {Journal of Statistical Physics}\ }\textbf
  {\bibinfo {volume} {144}},\ \bibinfo {pages} {759} (\bibinfo {year}
  {2011})}\BibitemShut {NoStop}%
\bibitem [{\citenamefont {Cicuta}\ \emph {et~al.}(2018)\citenamefont {Cicuta},
  \citenamefont {Krausser}, \citenamefont {Milkus},\ and\ \citenamefont
  {Zaccone}}]{Cicuta2018}%
  \BibitemOpen
  \bibfield  {author} {\bibinfo {author} {\bibfnamefont {G.~M.}\ \bibnamefont
  {Cicuta}}, \bibinfo {author} {\bibfnamefont {J.}~\bibnamefont {Krausser}},
  \bibinfo {author} {\bibfnamefont {R.}~\bibnamefont {Milkus}}, \ and\ \bibinfo
  {author} {\bibfnamefont {A.}~\bibnamefont {Zaccone}},\ }\href {\doibase
  10.1103/physreve.97.032113} {\bibfield  {journal} {\bibinfo  {journal}
  {Physical Review E}\ }\textbf {\bibinfo {volume} {97}} (\bibinfo {year}
  {2018}),\ 10.1103/physreve.97.032113}\BibitemShut {NoStop}%
\bibitem [{Note1()}]{Note1}%
  \BibitemOpen
  \bibinfo {note} {For $\delta z=0.1$ and $N=500$ and $1000$, we calculate
  $2\times 10^6$ matrices and for $N=2000$ and $4000$, we calculate $522240$
  and $261120$ matrices. For all other values, we calculate $10^6$
  matrices.}\BibitemShut {Stop}%
\bibitem [{\citenamefont {Xu}\ \emph {et~al.}(2010)\citenamefont {Xu},
  \citenamefont {Vitelli}, \citenamefont {Liu},\ and\ \citenamefont
  {Nagel}}]{Xu2010}%
  \BibitemOpen
  \bibfield  {author} {\bibinfo {author} {\bibfnamefont {N.}~\bibnamefont
  {Xu}}, \bibinfo {author} {\bibfnamefont {V.}~\bibnamefont {Vitelli}},
  \bibinfo {author} {\bibfnamefont {A.~J.}\ \bibnamefont {Liu}}, \ and\
  \bibinfo {author} {\bibfnamefont {S.~R.}\ \bibnamefont {Nagel}},\ }\href
  {https://doi.org/10.1209/0295-5075/90/56001} {\bibfield  {journal} {\bibinfo
  {journal} {{EPL} (Europhysics Letters)}\ }\textbf {\bibinfo {volume} {90}},\
  \bibinfo {pages} {56001} (\bibinfo {year} {2010})}\BibitemShut {NoStop}%
\bibitem [{\citenamefont {Ellenbroek}(2007)}]{Ellenbroek}%
  \BibitemOpen
  \bibfield  {author} {\bibinfo {author} {\bibfnamefont {W.~G.}\ \bibnamefont
  {Ellenbroek}},\ }\href@noop {} {\emph {\bibinfo {title} {Response of Granular
  Media near the Jamming Transition}}}\ (\bibinfo  {publisher} {Leiden
  Institute of Physics, Institute-Lorentz for Theoretical Physics, Faculty of
  Science, Leiden University},\ \bibinfo {year} {2007})\BibitemShut {NoStop}%
\bibitem [{\citenamefont {Lerner}\ and\ \citenamefont
  {Bouchbinder}(2018)}]{Lerner2017}%
  \BibitemOpen
  \bibfield  {author} {\bibinfo {author} {\bibfnamefont {E.}~\bibnamefont
  {Lerner}}\ and\ \bibinfo {author} {\bibfnamefont {E.}~\bibnamefont
  {Bouchbinder}},\ }\href {https://doi.org/10.1103/physreve.97.032140}
  {\bibfield  {journal} {\bibinfo  {journal} {Physical Review E}\ }\textbf
  {\bibinfo {volume} {97}} (\bibinfo {year} {2018})}\BibitemShut {NoStop}%
\bibitem [{\citenamefont {Benetti}\ \emph {et~al.}(2018)\citenamefont
  {Benetti}, \citenamefont {Parisi}, \citenamefont {Pietracaprina},\ and\
  \citenamefont {Sicuro}}]{Benetti2018}%
  \BibitemOpen
  \bibfield  {author} {\bibinfo {author} {\bibfnamefont {F.~P.~C.}\
  \bibnamefont {Benetti}}, \bibinfo {author} {\bibfnamefont {G.}~\bibnamefont
  {Parisi}}, \bibinfo {author} {\bibfnamefont {F.}~\bibnamefont
  {Pietracaprina}}, \ and\ \bibinfo {author} {\bibfnamefont {G.}~\bibnamefont
  {Sicuro}},\ }\href@noop {} {\enquote {\bibinfo {title} {Mean-field model for
  the density of states of jammed soft spheres},}\ } (\bibinfo {year} {2018}),\
  \Eprint {http://arxiv.org/abs/arXiv:1804.02705} {arXiv:1804.02705}
  \BibitemShut {NoStop}%
\bibitem [{\citenamefont {Dyson}(1953)}]{Dyson1953}%
  \BibitemOpen
  \bibfield  {author} {\bibinfo {author} {\bibfnamefont {F.~J.}\ \bibnamefont
  {Dyson}},\ }\href {\doibase 10.1103/physrev.92.1331} {\bibfield  {journal}
  {\bibinfo  {journal} {Physical Review}\ }\textbf {\bibinfo {volume} {92}},\
  \bibinfo {pages} {1331} (\bibinfo {year} {1953})}\BibitemShut {NoStop}%
\bibitem [{\citenamefont {Dean}(1964)}]{Dean1964}%
  \BibitemOpen
  \bibfield  {author} {\bibinfo {author} {\bibfnamefont {P.}~\bibnamefont
  {Dean}},\ }\href {\doibase 10.1088/0370-1328/84/5/310} {\bibfield  {journal}
  {\bibinfo  {journal} {Proceedings of the Physical Society}\ }\textbf
  {\bibinfo {volume} {84}},\ \bibinfo {pages} {727} (\bibinfo {year}
  {1964})}\BibitemShut {NoStop}%
\bibitem [{\citenamefont {Facoetti}\ \emph {et~al.}(2016)\citenamefont
  {Facoetti}, \citenamefont {Vivo},\ and\ \citenamefont
  {Biroli}}]{Facoetti2016a}%
  \BibitemOpen
  \bibfield  {author} {\bibinfo {author} {\bibfnamefont {D.}~\bibnamefont
  {Facoetti}}, \bibinfo {author} {\bibfnamefont {P.}~\bibnamefont {Vivo}}, \
  and\ \bibinfo {author} {\bibfnamefont {G.}~\bibnamefont {Biroli}},\ }\href
  {https://doi.org/10.1209/0295-5075/115/47003} {\bibfield  {journal} {\bibinfo
   {journal} {{EPL} (Europhysics Letters)}\ }\textbf {\bibinfo {volume}
  {115}},\ \bibinfo {pages} {47003} (\bibinfo {year} {2016})}\BibitemShut
  {NoStop}%
\end{thebibliography}%

\end{document}